\documentclass{aa}  

\usepackage{graphicx}
\usepackage[colorlinks=blue]{hyperref}
\usepackage{txfonts}
\usepackage{xcolor}
\begin{document} 

\title{Distinguishing between Flaring and Non-Flaring Active Regions: A Machine Learning Perspective}
\author{Soumitra Hazra$^{1,2, 3}$, Gopal Sardar $^{4, 5}$ and Partha Chowdhury $^6$}
\institute{
$^1$Université Paris-Saclay, CNRS, CEA, Astrophysique, Instrumentation et Modélisation de Paris-Saclay, 91191, Gif-sur-Yvette, France\\
$^2$Université Paris-Saclay, CNRS,  Institut d'astrophysique spatiale, 91405, Orsay, France\\
$^{3}$Center of Excellence and Space Sciences India, Indian Institute of Science Education and Research Kolkata, Mohanpur 741246, \\
West Bengal, India\\
\email{soumitra.hazra@cea.fr, soumitra.hazra@gmail.com}\\
$^{4}$Department of Physical Sciences, Indian Institute of Technology Jodhpur, Jodhpur 342011, India\\
$^{5}$Department of Physical Sciences, Indian Institute of Science Education and Research Kolkata, Mohanpur 741246, \\
West Bengal, India\\
\email{gopalsardar87@gmail.com}\\
$^6$Engineering Science Department, University of Calcutta, 87/1, College Street, Kolkata, India\\
\email{parthares@gmail.com}
   }
   \date{}
   
% \abstract{}{}{}{}{}  
% 5 {} token are mandatory
 
  \abstract
   {Large scale solar eruptions significantly impact space weather and damages space-based human infrastructures. It is necessary to predict large scale solar eruptions, which will enable us to protect our vulnerable infrastructures of modern society.}
   {We aim to investigate the difference between flaring and non-flaring active regions. We also want to investigate the forecasting probability of the solar flare.}
   {We use photospheric vector magnetogram data from Solar Dynamic Observatory's Helioseismic Magnetic Imager to study the time evolution of photospheric magnetic parameters on the solar surface. We build a database of flaring and non-flaring active region observed on the solar surface from the years 2010 to 2017. We train the machine learning algorithm by the time evolution of these active region parameters. Finally, we estimate the performance obtained from this machine learning algorithm.}
   {We find the strength of some magnetic parameters namely total unsigned magnetic flux, total unsigned magnetic helicity, total unsigned vertical current and total photospheric magnetic energy density in flaring active regions are much higher compared to the non-flaring ones. These magnetic parameters in the flaring active region are highly evolving and complex. We are able to obtain good forecasting capability with a relatively high value of true skill statistic (TSS). We also find that time evolution of total unsigned magnetic helicity and total unsigned magnetic flux have very high ability to distinguish flaring and non-flaring active regions.}
   {It is possible to distinguish flaring active region from the non-flaring one with good accuracy. We confirm that there is no single common parameter which can distinguish all flaring active regions from the non flaring one. However, time evolution of top few magnetic parameters namely total unsigned magnetic flux and total unsigned magnetic helicity have very high distinguishing capability.}


\titlerunning{Distinguishing between Flaring and Non-Flaring Active Regions}
\authorrunning{Hazra, Sardar \& Chowdhury}
   \maketitle
%
%________________________________________________________________

\section{Introduction}
Solar flare and coronal mass ejections are the two biggest explosions in the solar system. These two explosions release a huge amount of magnetic energy in the solar corona, creating disturbances in space weather. These two events directly impact the earth's atmosphere, causing geomagnetic disturbances. It is now well known that magnetic field structures in the sun are responsible for large scale eruptions. The study of magnetic fields in the sun is critical in understanding the energy build-up and release mechanism in solar flare and coronal mass ejection.

Solar flares and coronal mass ejections are believed to be a storage-and-release mechanism by which the non-potential magnetic field of the solar corona is released abruptly \citep{prie02, shib11}. It is also believed that complex magnetic structures on the solar surface are related to the onset of solar eruptions. Many studies have been performed to investigate the relationship between solar eruptions and photospheric magnetic parameters. Many active region parameters have been proposed to characterize the nonpotentiality of the magnetic field structures on the solar surface. Some of the well known nonpotentiality parameters are current helicity \citep{abra96, zhan99}, vertical electric current \citep{leka93}, horizontal gradient of the longitudinal magnetic field \citep{ziri93, tian02}, total photospheric magnetic free energy density \citep{wang96, metc05}, strong magnetic shear \citep{low77, kusa95}, reverse magnetic shear \citep{kusa04, vema12}, shear angle \citep{amba93}, twist parameter \citep{pevt94, hold04}  etc. Although individual case studies indicate a strong relationship between these nonpotentiality parameters and the flare productivity, it is till now not clear which property is common in all the eruptive active regions which will distinguish them from other non-eruptive active regions.

It is now well known that magnetic field structures on the solar surface change significantly with time. A detailed study of this photospheric magnetic field evolution may shed light on the energy build-up and release mechanism due to solar eruptions. The most frequently discussed mechanism for the change in photospheric magnetic field structure is flux emergence and cancellation \citep{livi89, spir02, sudo05, burt13}. Flux emergence and cancellation is found to play a significant role in some theories of the solar eruptions \citep{vanb89, amar10}. Flux cancellation is also one of the necessary conditions for the formation of solar filaments \citep{marti85, gaiz97, mart01}. Solar filaments are believed to be one of the major precursors for solar eruptions \citep{sinh19}. One can also predict the possible orientation of the ejected magnetic field due to solar eruptions by studying the hemispheric preference of the filament chirality \citep{marti94, hazr18}.  In summary, the time evolution of photospheric magnetic field parameters plays an important role in the onset phase of the solar flare. However, given a large amount of solar data, it is almost impossible to analyze every solar eruptive event manually. We must have to build some reliable automated method which can analyze the eruptive active regions and distinguish them from other non-eruptive active regions.

In recent times, machine learning appears as a promising automated candidate for reliable forecasting of solar eruptive events \citep{ahme13, bobr15, bobr16, nish17, hamd17, ma17, boub18, flor18, ince18}. Machine learning is also used to identify the common parameter which is most important to distinguish the eruptive active region from other non-eruptive active regions. \cite{dhur19} used machine learning to find out the critical criteria at the onset phase which can lead to a solar flare. Different types of data sets are used for the purpose of predicting eruptive events using machine learning. \cite{yu09} and \cite{yuan10} used line of sight magnetogram data obtained from Michelson Doppler Imager (MDI) for flare prediction. \cite{agga18} used filament metadata for the prediction of eruptive events. However, most of the studies used vector magnetogram
data obtained from the Helioseismic Magnetic Imager (HMI) onboard Solar Dynamics Observatory (SDO) for the purpose of flare prediction. Different machine learning classifiers have been used for the solar flare prediction. Some studies also used the time series of the magnetic field data obtained from HMI for flare prediction \citep{hamd17, ma17}.

In this paper, we aim to investigate the importance of the time evolution of magnetic parameters in terms of flare forecasting. We find that there is a significant difference between the eruptive and non-eruptive active region in terms of both strength and time evolution of photospheric magnetic parameters. We also try to find out the common magnetic parameter which will clearly separate the eruptive active regions from the non-eruptive ones. For this purpose, we train the machine learning algorithm using the time evolution of photospheric magnetic parameters. We are able to predict solar flare quite well. We find that total unsigned magnetic helicity and total unsigned magnetic flux have higher distinguishing capability compared to other photospheric magnetic parameters. 

Section 2 describes the details of the data used in this study. We present a detailed manual comparison study between eruptive and non-eruptive active regions in terms of the time evolution of magnetic parameters in Section 3. We present a comparison study between eruptive and non-eruptive active regions using the machine learning algorithm in Section 4. We also describe the details about the machine learning algorithm and its performance in Section 4. Finally, we present the summary and conclusions of our study in Section 5.

\section{Data}
\subsection{Data for Active Regions}
Helioseismic Magnetic Imager (HMI), an instrument onboard Solar Dynamics Observatory (SDO) spacecraft, provides us continuous full-disk photospheric magnetic field data \citep{sche12, scho12}. HMI team developed an automated method that detects active region patches from the full disk vector magnetogram data and provides us a derivative data named, Space-weather HMI Active Region Patches (SHARP) \citep{bobr14}. The automatic detection algorithm operates on the line of sight magnetic field image and creates a smooth bounding curve, called bitmap centered at the flux weighted centroid. The HMI Stokes I, Q, U, V data was inverted within the smooth bounding curve by Very Fast Inversion of the Stokes Vector (VFISV) code, which is based on the Middle-Eddington model of the solar atmosphere. The $180^{\circ}$ ambiguity in the transverse component of the magnetic field was corrected using the minimum-energy algorithm \citep{metc94, crou09}. The inverted and disambiguated magnetic vector field data has been remapped to a Lambert Cylindrical Equal-Area projection which gives us decomposed Bx, By and Bz data. JSOC provides us this decomposed data. We have downloaded this decomposed data from the JSOC webpage.  We have calculated 17 active region magnetic field parameters in every 12 minutes from this SHARP data. These parameters are listed with keywords and formula in a table (see below). Please note that we follow the same procedure to calculate active region magnetic field parameters as defined in \cite{bobr15}. We consider the pixels which are within bitmap and above a high confidence disambiguation threshold (coded value is greater than 60) for our magnetic parameter calculation. We use a finite difference method for calculation of computational derivative needed for parameter calculation. We use Green's function technique with a monopole depth of 0.00001 pixels for calculation of potential magnetic field which is necessary for the calculation of the total photospheric magnetic free energy density. We neglected active regions near the limb, where it is difficult to see magnetic structures due to the projection effect. Calculated magnetic field parameter data is also not reliable near the limb. Thus we only consider the data for our study which is within $\pm 70^{\circ}$ from the disc center. We note that data for all of these magnetic parameters are available in the SHARP header. SHARP data products from SDO HMI can be found at \href{http://jsoc.stanford.edu/ajax/lookdata.html?ds=hmi.sharp_cea_720s}{\textcolor{blue}{jsoc.stanford.edu}} \citep{sche12}.

\subsection{Data for Solar Flare}
We consider the solar flare for our study based on the peak X-ray flux observed by GOES X-ray satellites. When Goes satellite detects a flare, it generally reports to the flare catalog. Then the flare is paired with its parent active region. Generally, five types of flares namely A, B, C, M, and X are observed by GOES satellites. While X and M class are high-intensity flares (intensity greater than $10^{-5}$~Wm$^{-2}$); other A, B, C flares are less intensive ones. For our study, we only consider X and M class flares as a flare. We also only consider the flares for our study which are within $\pm 70^{\circ}$ of the central meridian and if there is also an associated parent active region. 
\begin{table*}[h]
\begin{center}
 \begin{tabular}{|r r r r| }
 \hline
 Keyword & Description & Unit & Formula \\ [0.50ex] 
 \hline
 TOTUSJH & Total unsigned current helicity & G$^2$/m & $H_{c_{total}} \propto \sum |B_z.J_z|$  \\ 
 
TOTPOT & Total photospheric magnetic free energy density &ergs/cm& $\rho_{tot} \propto \sum (\vec{B}^{obs} - \vec{B}^{pot})^2 dA $   \\
 
 TOTUSJZ & Total unsigned vertical current & Amperes & $J_{z,total} \propto \sum |J_z| dA$  \\ 
 
 SVANCPP & Sum of the modulus of the net current per polarity & Amperes & $J_{z,sum} \propto |\sum \limits^{B_z^+}  J_z dA| + |\sum \limits^{B_z^-}  J_z dA|$  \\ 
 
  ABSNZH & Absolute value of the net current helicity & G$^2$/m & $H_{c_{abs}} \propto |\sum B_z.J_z|$  \\ 
 
 USFLUX  & Total unsigned flux & Maxwell & $\Phi  \propto \sum |B_z| dA$  \\ 
 
 MEANPOT & Mean photospheric magnetic free energy density & ergs/cm$^3$ & $\bar{\rho} \propto \frac{1}{N} \sum (\vec{B}^{obs} - \vec{B}^{pot})^2$  \\ 
 
 MEANGAM  & Mean angle of field from radial & Degrees &  $\bar{\gamma}  \propto \frac{1}{N} \sum arctan(\frac{B_h}{B_z})$  \\ 
 
 MEANSHR  & Mean shear angle & Degrees & $\bar{\Gamma}  \propto \frac{1}{N}  \sum arccos (\frac{\vec{B}^{obs}.\vec{B}^{pot}}{|B^{obs}|.|B^{pot}|})$  \\ 
 
  SHRGT45  & Fraction of area with shear $> 45^{\circ}$ & m$^2$ & $Area ~ with~shear > 45^{\circ} / total~area$  \\ 
 
  AREA$\_$ACR  & Area of strong field pixels in a active region & m$^2$ & $Area  =  \sum Pixels$  \\ 
 
  MEANGBT  & Mean gradient of total field & G/Mm &$\overline{|\nabla B_{tot}|}  = \frac{1}{N} \sum \sqrt{(\frac{\partial B}{\partial x})^2 + (\frac{\partial B}{\partial y})^2} $  \\ 
 
 MEANGBZ  & Mean gradient of vertical field & G/Mm & $\overline{|\nabla B_{z}|}  =  \frac{1}{N} \sum \sqrt{(\frac{\partial B_z}{\partial x})^2 + (\frac{\partial B_z}{\partial y})^2} $     \\ 
 
  MEANGBH  & Mean gradient of horizontal field & G/Mm & $\overline{|\nabla B_{h}|}  =  \frac{1}{N} \sum \sqrt{(\frac{\partial B_h}{\partial x})^2 + (\frac{\partial B_h}{\partial y})^2} $   \\ 
 
MEANJZH  & Mean current helicity & G$^2$/m & $\bar{H_c}  \propto \frac{1}{N} \sum B_z J_z$  \\ 
 
MEANJZD  & Mean vertical current density & mA/m$^2$ & $J_z  \propto \frac{1}{N} \sum (\frac{\partial B_y}{\partial x} - \frac{\partial B_x}{\partial y})$  \\ 
 
 MEANALP  & Mean characteristic twist parameter, $\alpha$  & 1/Mm & $\alpha_{total}  \propto \frac{\sum J_z B_z}{\sum B_z^2}$  \\ 
 \hline 
\end{tabular}
\end{center}
\end{table*}

 \begin{figure*}
        \centering
        \includegraphics[width=18 cm]{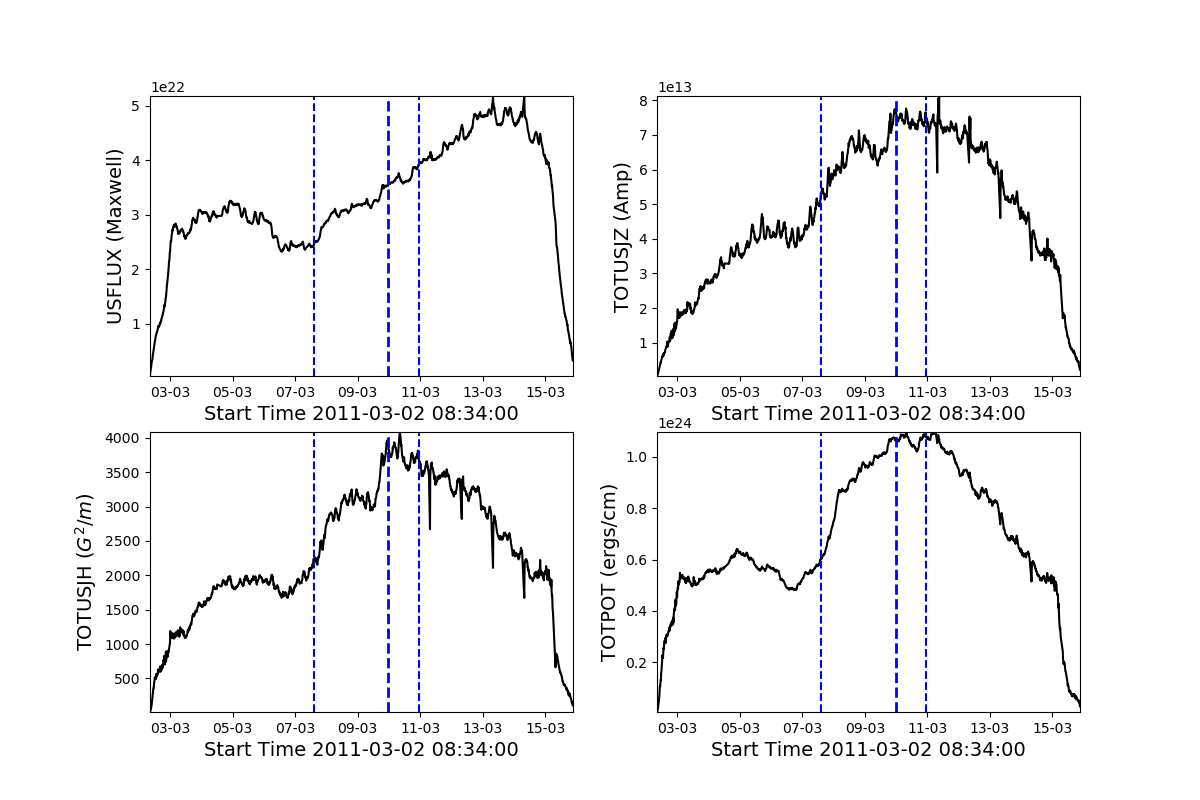}
        \caption{Time evolution of total unsigned magnetic flux (USFLUX), total unsigned vertical current (TOTUSJZ), total unsigned current helicity (TOTUSJH) and total photospheric magnetic free energy density (TOTPOT) for activee region NOAA 11166 (SHARP 401). Thick dashed blue vertical line corresponds to X class solar flare and thin dashed blue vertical line correspond to M class solar flare.}
        \label{fig:1}
\end{figure*}

  \begin{figure*}
        \centering
        \includegraphics[width=18 cm]{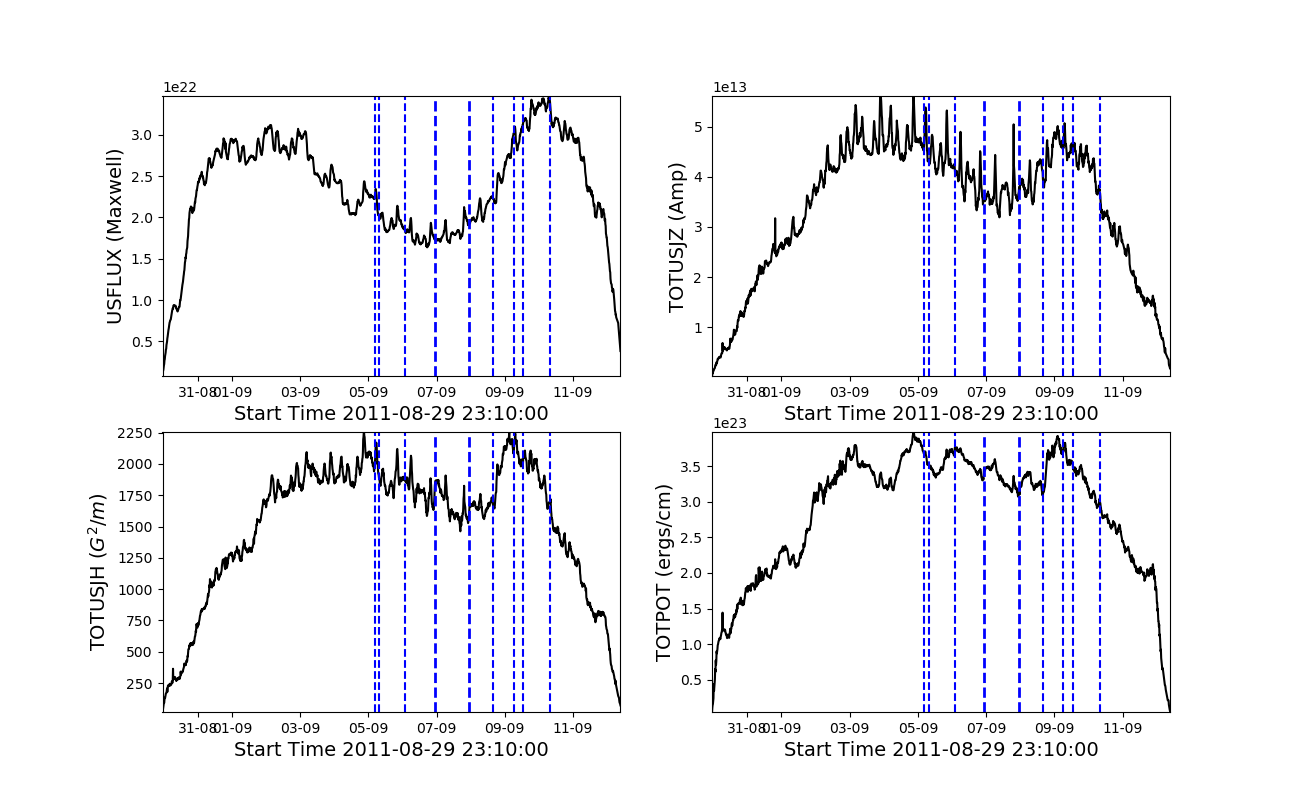}
        \caption{Time evolution of total unsigned magnetic flux (USFLUX), total unsigned vertical current (TOTUSJZ), total unsigned current helicity (TOTUSJH) and total photospheric magnetic free energy density (TOTPOT) for activee region NOAA 11283 (SHARP 833). Thick dashed blue vertical line corresponds to X class solar flare and thin dashed blue vertical line correspond to M class solar flare.}
        \label{fig:2}
\end{figure*}

  \begin{figure*}
        \centering
        \includegraphics[width=18 cm]{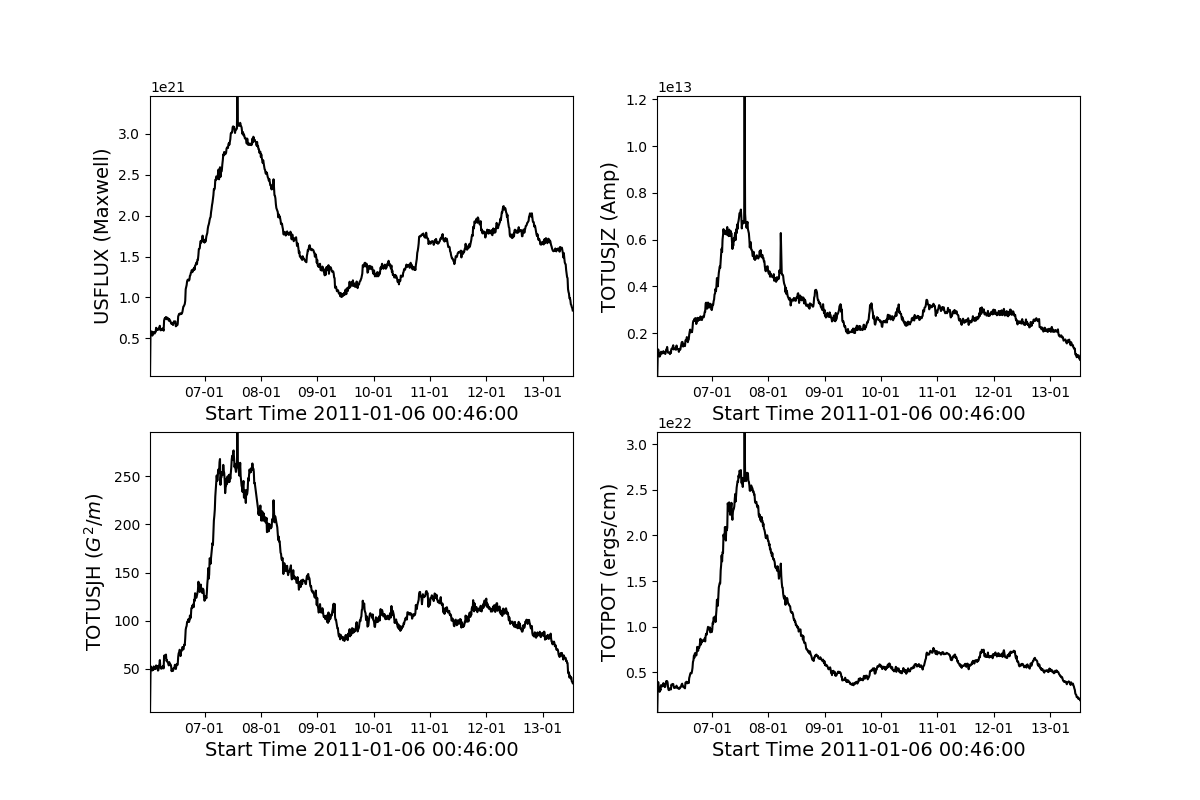}
        \caption{Time evolution of total unsigned magnetic flux (USFLUX), total unsigned vertical current (TOTUSJZ), total unsigned current helicity (TOTUSJH) and total photospheric magnetic free energy density (TOTPOT) for activee region NOAA 11143 (SHARP 335). This active region does not produce any solar flare during its transit over visible solar disk.}
        \label{fig:2}
\end{figure*}
\section{Comparing Flaring and Non-Flaring Regions}
Active regions NOAA 11166 (SHARP 401), NOAA 11283 (SHARP 833) and NOAA 11143 (SHARP 335) were chosen for the comparison study between flaring and the non-flaring active region. All of these active regions transit over the visible solar disk for a long duration. AR 11166 produces one X class and two M class solar flares during the passage over the visible solar disk, while AR 11283 produces two X class and five M class solar flares. In contrast, AR 1143 produces no flare during its transit. One may pose the question- why these three active regions behave so differently during their transit over the solar disk? is it possible to distinguish flaring and non-flaring active regions?

It is now well known that the different magnetic nature of the active regions is responsible for different behaviors. Here, we study the temporal evolution of photospheric active region magnetic parameters to get an idea about the difference between flaring and non-flaring active regions. Figures 1, 2, and 3 show the temporal evolution of four magnetic parameters--the total unsigned magnetic flux ($\Phi_{tot}$), total unsigned current helicity ($h_{c,tot}$), total unsigned vertical current ($J_{z,tot}$) and the proxy of total photospheric magnetic free energy density ($\rho_{tot}$). All these four parameters have much a higher value in the case of the flaring active regions (AR 11166 and 11283) compared to the non-flaring active region (AR 11143). All four magnetic parameters also show significant evolution. Total unsigned magnetic flux for AR 11166 although decreases before the first large scale flare but increases for the other two flare (Fig.~1). The other three magnetic parameters for AR 11166 show an increasing trend before the first flare. Total unsigned magnetic flux for AR 11283 shows a significant decreasing trend before the first flare and increasing trend later (Fig.~2). Total unsigned magnetic helicity for both flaring active regions (AR 11166 and 11283) shows an increasing trend before the first flare and both the active region starts the flaring activity when the value of magnetic helicity is sufficiently high. Another interesting point is that once an active region starts flaring, it keeps on flaring. All four magnetic parameters also show significant evolution for non-flaring AR 11283 but have a much lower value compared to the other two flaring active regions (see Fig.~3). We also note that there is a high signal-to-noise ratio in the data near the solar limb, thus in our time evolution study, the value of magnetic parameters are not reliable during the start and end time (active regions are near the limb).

Change in the total unsigned magnetic flux during the active region transit is mostly due to flux cancellation and emergence on the solar surface. The disappearance of the magnetic flux is always observed when the magnetic flux of one particular polarity encounters the flux fragments of the opposite polarity. Some previous studies indicate that flux cancellation plays an important role in triggering solar eruptions \citep{vanb89, amar10}. The total unsigned magnetic flux of the active region and the magnetic flux near the polarity inversion line (R-value) is also found to be correlated well with the flaring activity and the coronal X-ray luminosity \citep{schr07, leka07, barn08, burt13, hazr15}. The emergence of the new magnetic flux is also a well-observed phenomenon and believed to be one of the mechanisms for the formation of the current sheet \citep{tur76,wang93, sudo05}. Our results point out that total unsigned magnetic flux is considerably higher in flaring active regions compared to non-flaring active regions. 

We also found that flaring active regions are magnetically more complex compared to non-flaring active regions. One can characterize the magnetic complexity in an active region by different magnetic parameters, namely vertical electric current, magnetic helicity, twist, shear angle, photospheric magnetic free energy density, etc \citep{abra96, metc05, pevt94, park08}. Magnetic helicity, a measure of twist, shear, and inter-linkage of magnetic field lines, is a conserved quantity in an ideal MHD scenario \citep{berg84}. Change in the magnetic helicity reflects the deviation from the ideal MHD scenario, indicating the evolution of magnetic complexity inside the active regions. We find significantly higher magnetic helicity and excess magnetic free energy in the flaring active region compared to the non-flaring region. Recent theoretical and observational studies also suggest that the injection of magnetic helicity of both the same and opposite sign to system's global helicity can trigger the solar flare \citep{kusa02, moon02, park08, park12, park13}. \cite{kusa03} also developed a theoretical model of solar flare based on the annihilation of magnetic helicity. Our result indicates both the accumulation and annihilation of magnetic helicity before the onset of a solar flare.

In summary, our result highlights the importance of time evolution of magnetic parameters in distinguishing flaring and non-flaring active regions. However, one may pose the question -- which magnetic parameter is more important? It is very difficult to tell. In reality, there is a large number of active regions appeared on the solar surface within a few days. It is difficult to analyze all active regions manually to predict the probability of that active region eruption. We have to develop some automated method which will help us to predict whether an active region will flare or not. 

\section{Comparing Flaring and Non-Flaring Regions using Machine Learning}
In the previous section, we discussed the differences between flaring and non-flaring active regions based on the time evolution of magnetic parameters. In this section, we want to distinguish them based on the automated machine learning method. Machine learning is a branch of artificial intelligence which provides computer the ability to learn automatically and improve from the past experience without being explicitly programmed. Two types of learning namely, unsupervised learning and supervised learning exist in the machine learning literature. In an unsupervised learning scenario, we do not have to supervise the model. We will allow the computer model to work on its own to figure out the information. This learning technique mainly works with unlabelled data. In the supervised machine learning scenario, we have to train the computer model based on known well-labeled data. As we already have a well-known database of flaring and non-flaring active regions, we have used supervised learning techniques for our problem. In the supervised machine learning scenario, first, we have to train the computer-based on known flaring and non-flaring active region data. After the training, the computer will be able to tell us the possibility of active region eruption.

 \begin{figure*}
        \centering
        \includegraphics[width=4 cm ,angle =-90  ]{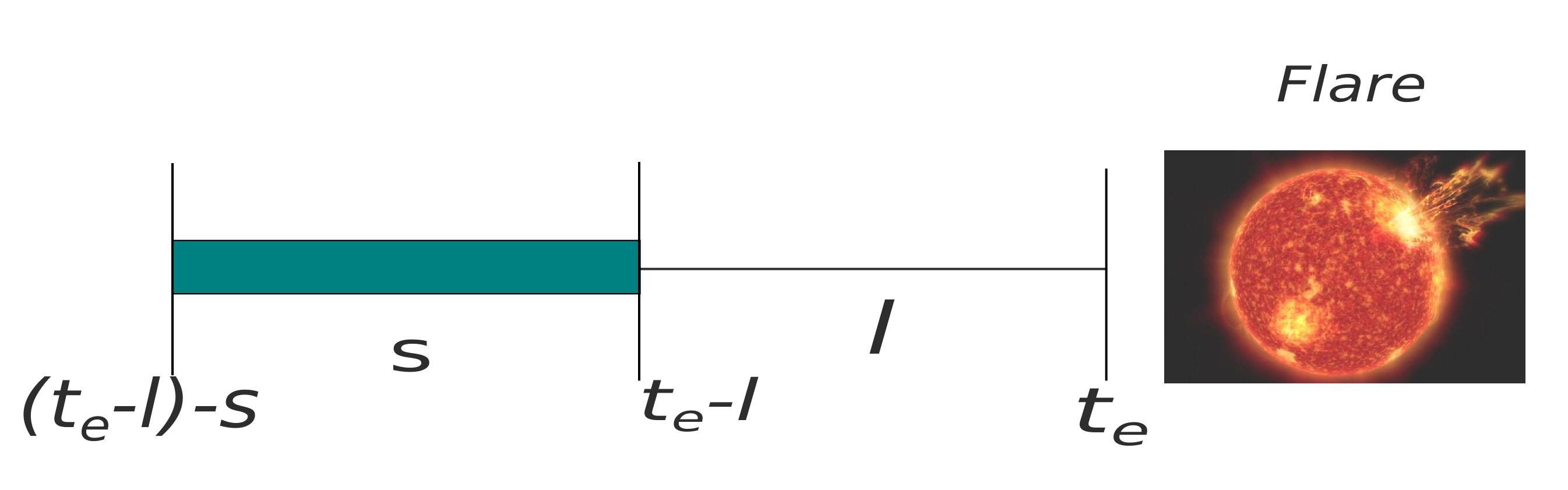}
        \caption{Data selection criteria for eruptive active regions in terms of flare generating time ($t_e$), loopback time ($l$), and span time ($s$). We use the time evolution of active region parameters during span time ($s$) for training the machine learning algorithm. loopback time is also the forecasting window. }
        \label{fig:4}
\end{figure*}
 
 \subsubsection{Data Preparation}
To train the supervised machine learning algorithms, we have to first define the positive and negative class. We follow the same definition as prescribed by \cite{ma17} and \cite{dhur19} for this purpose. The active region which produces at least one X or M class solar flare during its transit over visible solar disk belongs to the positive class and the active region which does not produce any X or M class flare during its transit belongs to the negative class.

Most of the previous flare prediction studies consider only magnetic parameters which are 24 hours before the flare time. They did not consider any time evolution of magnetic parameters for their prediction purpose. However, most of the theoretical models suggest a change in magnetic parameters before the solar flare. The standard Flare model indicate that flux cancellation near the polarity inversion line is an important determinant for solar flare \citep{vanb89}. Thus it is necessary to include the basic essence of the time evolution of magnetic parameters in the training purpose.

In this study, we consider the time evolution of magnetic parameters during a time window, named span, for the training purpose. The span time window is always before the loopback time. The loopback is the time window before the occurrence of the solar flare. Figure 4 shows a graphical representation of this selection. Let assume, we want to predict solar flare before 24 hours of the occurrence, and for this purpose, we consider the 12 hours of the time evolution of magnetic parameters which is 24 hours before the flare occurrence. In this situation, loopback is 24 hours and span is 12 hours.  As we consider 17 magnetic parameters for our study, so we get 17 time series of magnetic parameter evolution for the training purpose. We represent each time series of magnetic parameter evolution by seven statistical parameters associated with the time series, namely, mean, median, skewness, kurtosis, standard deviation, first and third quartile. For a time series named $T =[x_1, x_2, x_3,..., x_n]$, statistical summary parameters are defined in the following way:
\begin{equation}
 Mean~~(\mu) =   \frac{1}{N} \sum_{i=1}^{N} x_i
\end{equation}
\begin{equation}
 Std.~Deviation~(\sigma) = \sqrt{\frac{ \sum_{i=1}^{n} (x_i - \mu(T))^2}{N}}
\end{equation}
\begin{equation}
Skewness = \frac{ \sum_{i=1}^{n} (x_i - \mu(T))^3}{N \sigma(T)^3}
\end{equation}
\begin{equation}
 Kurtosis = \frac{ \sum_{i=1}^{n} (x_i - \mu(T))^4}{N \sigma(T)^4}  . 
\end{equation}

Median is the middle number of the ascending time series T, the first quartile is the middle value between the median and the smallest number of the time series T and third quartile is the middle value between the median and the largest number of the time series T. Standard deviation ($\sigma$) basically represents the dispersion around the mean. Skewness and kurtosis is basically a statistical measure to describe the distribution. While skewness is the measure of the symmetry of the dataset; kurtosis tells us how the tails of the distribution differ from the tails of the normal distribution. We consider 17 magnetic parameters for our study. So, we have a time series for every seventeen magnetic parameters. As we represent each of the time series by 7 statistical parameters, so we have 119 entries in the resultant matrix. 

\subsection{Different Supervised Machine Learning Techniques}
 There are different supervised machine learning classifiers in the literature. These supervised machine learning algorithms are used for training. Some of the well known supervised machine learning classifiers are- logistic regression, decision tree, KNN, Naive Bayes, support vector machine, multilayer perceptron, random forest, etc. In the machine learning literature, every problem is unique. We do not know what algorithms to use, whether the problem can be modeled effectively. A baseline model is the simplest possible prediction model. This baseline model result will tell us whether the use of more advanced algorithms are adding any value in the result or not. There is no need for a complex, advanced machine learning algorithm for a particular problem if a simple baseline algorithm can do the same. In this study, we use the Logistic Regression classifier as a baseline model and compare the baseline result with the results obtained from some other complex machine learning algorithms, namely, support vector machine, and multilayer perceptron.

{\bf Baseline Model: Logistic Regression}\\
Logistic Regression (LR) is one of the simplest and commonly used machine learning algorithms for the binary classification problem. It is easy to implement, easily interpretable, efficient and does not require high computation power; thus it can be used as a baseline model for the binary classification problem. This model estimates the probability of an event occurrence by fitting data to a logistic function. The equation used for logistic regression is:\\
$log(\frac{p}{1-p})= \alpha_0 + \alpha_1 x_1 + .... +\alpha_n x_n$\\
where p is the probability of the event occurrence. $x_1, x_2, .... x_n$ is the number of independent variables. $\frac{p}{1-p}$ is known as the odd ratio. If the odd ratio is positive, then the probability of event occurrence is more than 50 \%. One of the major drawbacks of the algorithm is the assumption of linearity between the dependent and independent variable. This algorithm separates the classes by constructing a linear decision boundary between them. It is basically a linear classifier. It does not perform well if the classes are not linearly separable. We use {\it lbfgs} solver for our logistic regression classifier.

{\bf Support Vector Machine:} Support Vector Machine (SVM) is a classification algorithm that separates the data of two classes by finding a line (in 1-D) or a hyperplane (in higher dimensions) between two classes. In the SVM algorithm, the points near the line or the hyperplane are called support vector and the distance between the support vector and the line or the hyperplane is called margin. This algorithm tries to find hyperplane or line by maximizing the margin. SVM is highly suitable for linear classification problems. However, SVM can also solve nonlinear classification problems by moving lower-dimensional space to higher-dimensional space such that we can find the separating hyperplane in the higher dimension. These transformations are known as kernel trick.

Let assume we have N training points where each input $x_i$ has $D$ attributes and belongs to any of the two classes $y_i= +1$ or $-1$. In most of the cases, different classes are not-fully linearly solvable. In this situation, people generally use the soft margin SVM algorithm where the concept of a slack variable and the idea of a trade-off between the minimization of misclassification rate and the maximization of margin is introduced. One can describe the hyperplane by the equation $w_i x_i + b -1 =0$ where $w$ is the normal to the hyperplane and $b/||w||$ is the normal distance from origin to the hyperplane. In the SVM scenario, $1/||w||$ is the margin. In the soft margin SVM algorithm, we have to select the variable $b$ and $w$ in a way so that we can describe our training data by:\\

    $x_i.w + b \geq +1 - \psi_i$   for $y_i = +1$
    
    $x_i.w + b \leq -1 + \psi_i$   for $y_i = -1$
    
where, $\psi_i \geq 0$ is the slack variable. We can combine these two equation into a single equation:

$y_i (x_i.w + b) -1 + \psi_i \geq 0$

In the soft margin SVM algorithm, one has to maximize the margin and also have to reduce the misclassification rate. This can be done by minimizing an objective function subject to the previous condition. In a more general way, this problem can be defined as:
\begin{equation}
    min\left(\frac{1}{2} w^Tw + C \sum_{i=1}^{L} \psi_i\right)
\end{equation}
such that
$y_i (w^T \Phi(x_i) + b) -1 + \psi_i \geq 0$. Parameter C controls the trade-off between the size of the margin and the slack variable. This parameter is also known as the regularization parameter. $\Phi$ is the function that maps the input data into higher dimensional space, also known as the kernel function. It is sometimes tricky to find an appropriate kernel for a particular problem. We do not know whether our problem is linearly separable or not linearly separable. We have used kernel trick for our problem. It has been previously shown that radial basis function kernel projects vectors into infinite-dimensional space. Motivated by this fact, we have used a radial basis function kernel for our study. Choice of the rbf kernel ensures that our support vector machine algorithm will separate the classes by constructing a non-linear decision boundary.

{\bf Multilayer Perceptron:} Multilayer perceptron (MLP) uses the concept of a neural network to predict the output based on some input features. Perceptron is a linear classifier which separates the input into two classes by a straight line. The output of a perceptron depends on the input $i.e.,$ the feature vector (x) which is multiplied by weight w and added to a bias b (simply, output $=w.x + b$). The final prediction will be after passing the output of the perceptron through a non-linear activation function. 

Multilayer Perceptron is basically a deep neural network. It consists of an input layer where the feature vector is fed, an output layer for making the prediction about the input and an arbitrary number of hidden layers in between the input and output layer. Each neuron in the multilayer perceptron is connected with all other neurons of the previous layer. Neuron is basically a processing unit where inputs are summed using weights and result is passed via an activation function. In summary, the output of each basic processing unit (neuron) is:
\begin{equation}
    y= \Phi\left(\sum{w_i x_i +b}\right)= \Phi(w^{T}x+b)
\end{equation}

where x denotes the vector of inputs, w is the vector of weights, b represents the bias and $\Phi$ is the activation function. We use the RELU activation function for each neuron in the hidden layers and in the output layer, we use a sigmoid activation function.

Training the MLP algorithm involves the adjustment of bias and weights to minimize the error in the output. This is achieved by using forward and backward propagation. MLP is basically a feed-forward network which involves constant backward and forward propagation until we achieve the desired result.

{\it Forward Propagation:} In this method, we move the signal from the input layer to the output layer via hidden layers. We measure the output or decision of the output layer with respect to the ground truth label. This measurement is also known as error.

{\it Backward Propagation:} In this process, we backpropagate the partial derivative of the error with respect to weights and bias from the output layer. A stochastic gradient descent algorithm is used to adjust the weights and bias in this process.

The multilayer perceptron model has some advantage. A very complex model can be trained by the MLP model and feature engineering is not required before training. However, it is difficult to explain the MLP model simply and parameterization is also complex. This model also needs more training data. 
\subsubsection{Performance Measure and Class Imbalance Problem}
  We will get a confusion matrix as a result of our classification algorithm- which consists of four entries, namely, TP, TN, FP, and FN. Here TP (True Positive) are the cases where positively labeled samples are correctly predicted as positive, TN (True Negative) are the cases where positively labeled samples are wrongly predicted as negative, FP (False Positive) are the cases where negatively labeled samples are predicted as positive and FN (False Negative) are the cases where negatively labeled samples are correctly predicted as negative. Accuracy in the classification problem is defined as the number of correct predictions made by the model over the total number of predictions made:
  \begin{equation}
      Accuracy= \frac{TP + TN}{TP + FP + TN + FN}
  \end{equation}
  Accuracy is a good performance measure when the data set is balanced. Some other performance measures are Precision ($TP/(TP+FP)$), Recall ($TP/(TP + FN)$) and the F-score (harmonic mean of precision and recall).\\
  
  As solar active regions do not have flare/eruption most of the time, thus non-flaring active regions are much more compared to the number of flaring active regions. There is a huge imbalance between the number of flaring and non-flaring active regions. This problem is known as the class imbalance problem. In this case, accuracy will be very high if the model predicts almost all active regions as non-flaring (as the number of non-flaring regions is very high). However, we aim to predict the flaring active regions which are rare. Thus, accuracy is not a good performance measure for the class imbalance problem. Later, some other performance measures, namely, Heidke Skill Score ($HSS1$ and $HSS_2$) and Gilbert Score (GS) have been used \citep{barn08, maso10, azim19}:
  \begin{equation}
      HSS_1= \frac{TP + TN -N}{P}
  \end{equation}
  
  \begin{equation}
      HSS_2= \frac{ 2 \times [(TP \times TN) - (FN \times FP)]}{P \times (FN + TN) + (TP + FP) \times N}
  \end{equation}
  \begin{equation}
      GS= \frac{TP \times (P + N) - (TP + FP) \times (TP + FN)}{(TP + FP + FN) \times (P + N) - (TP + FP) \times (TP + FN)}
  \end{equation}
  
  where, $P$ and $N$ are the total number of actual positive and negative samples. $HSS_1$ measures the improvement of the prediction over all negative predictions, while $HSS_2$ measures the same but over a random forecast. Gilbert Score (GS) measures the number of true positive (TP) obtained by chance.
  
  However, all these measures $HSS_1$, $HSS_2$ and $GS$ till have some dependency on the ratio of class imbalance. To alleviate this problem, \cite{bloom12} introduced a new performance measure, namely True Skill Statistic (TSS) which is independent of the class imbalance ratio. True skill statistic is defined as:
  \begin{equation}
      TSS = \frac{TP}{TP + FN} - \frac{FP}{TN + FP}.
  \end{equation}
 Value of TSS varies from $-1$ to $+1$ where perfect correct prediction scores $+1$, always wrong prediction scores $-1$ and random prediction scores zero. Flare prediction is a highly imbalanced problem. TSS is the most meaningful measure in the case of flare prediction scenario as it does not depend on the ratio of class imbalance.
  \begin{table*}[t]
\caption[]{Flare prediction capabilities obtained from our baseline model Logistic Regression for five different data sets. First three data sets namely loop24span12, loop24span24, loop24span0 correspond to the data sets with same forecasting window of 24 hour but different span time of 12 hour, 48 hour and zero hour respectively. While last two data sets namely loop12span12 and loop48span12 correspond to the data sets with same span time of 12 hour but different forecasting window of 12 hour and 48 hour respectively. }
 \begin{center}\begin{tabular}{c|c|c|c|c|c|c|}
 \hline
 \multicolumn{7}{c} {Results from Our Baseline Model: Logistic Regression}\\
 \hline
   &  & loop24span12 & loop24span24& loop24span0 & loop12span12& loop48span12 \\
  \hline
Considering   & Accuracy & 0.95 $\pm$ 0.023 & 0.94 $\pm$ 0.007 &0.94 $\pm$ 0.008 & 0.96 $\pm$ 0.006 & 0.92 $\pm$ 0.010\\
all  & Precision (Positive) & 0.84 $\pm$ 0.023 &  0.84 $\pm$ 0.019 & 0.79 $\pm$ 0.026 & 0.87 $\pm$ 0.021 & 0.77 $\pm$ 0.025 \\
magnetic  & Precision (Negative) &0.98 $\pm$ 0.005 & 0.98 $\pm$ 0.005 & 0.98 $\pm$ 0.005 & 0.97 $\pm$ 0.003 & 0.97 $\pm$ 0.006\\
parameters  & Recall (Positive) & 0.93 $\pm$ 0.018& 0.92 $\pm$ 0.016 & 0.93 $\pm$ 0.019 & 0.96 $\pm$ 0.014 & 0.89 $\pm$ 0.021\\
& Recall (Negative) &0.95 $\pm$ 0.008 & 0.96 $\pm$ 0.006 & 0.94 $\pm$ 0.010 & 0.97 $\pm$ 0.004 & 0.92 $\pm$ 0.011 \\
& F$_1$-score (Positive) &0.88 $\pm$ 0.014 &0.88 $\pm$ 0.013 &0.85 $\pm$ 0.017 & 0.91 $\pm$ 0.014 & 0.83 $\pm$ 0.018 \\
& F$_1$-score (Negative) & 0.97 $\pm$ 0.004 & 0.97 $\pm$ 0.004 & 0.96 $\pm$ 0.005 & 0.97 $\pm$ 0.004 & 0.94 $\pm$ 0.007\\
& HSS$_1$ &0.93 $\pm$ 0.008 & 0.94 $\pm$ 0.007 & 0.92 $\pm$ 0.010 & 0.96 $\pm$ 0.007 & 0.89 $\pm$ 0.012\\
& HSS$_2$ &0.84 $\pm$ 0.018 & 0.84 $\pm$ 0.016 & 0.82 $\pm$ 0.023 &0.89 $\pm$ 0.018 & 0.77 $\pm$ 0.024 \\
& GS & 0.73 $\pm$ 0.027 &  0.73 $\pm$ 0.025 & 0.69 $\pm$ 0.032 &0.81 $\pm$ 0.029 & 0.64 $\pm$ 0.032 \\
& TSS &0.88 $\pm$ 0.017 & 0.87 $\pm$ 0.016 & 0.87 $\pm$ 0.021 & 0.92 $\pm$ 0.015 & 0.82 $\pm$ 0.023\\
\hline
Considering   & Accuracy & 0.94 $\pm$ 0.006 & 0.93 $\pm$ 0.007 &0.94 $\pm$ 0.006 & 0.96 $\pm$ 0.005 & 0.91 $\pm$ 0.009\\
only top  & Precision (Positive) & 0.83 $\pm$ 0.023 &  0.81 $\pm$ 0.022 & 0.83 $\pm$ 0.022 & 0.85 $\pm$ 0.020 & 0.78 $\pm$ 0.025 \\
five magnetic  & Precision (Negative) &0.97 $\pm$ 0.006 & 0.97 $\pm$ 0.006 & 0.96 $\pm$ 0.005 & 0.98 $\pm$ 0.005 & 0.95 $\pm$ 0.008\\
parameters  & Recall (Positive) & 0.88 $\pm$ 0.023& 0.87 $\pm$ 0.024 & 0.86 $\pm$ 0.023 & 0.91 $\pm$ 0.021 & 0.84 $\pm$ 0.027\\
& Recall (Negative) &0.95 $\pm$ 0.008 & 0.95 $\pm$ 0.007 & 0.96 $\pm$ 0.004 & 0.96 $\pm$ 0.006 & 0.93 $\pm$ 0.011 \\
& F$_1$-score (Positive) &0.85 $\pm$ 0.015 &0.84 $\pm$ 0.017 &0.84 $\pm$ 0.014 & 0.88 $\pm$ 0.014 & 0.81 $\pm$ 0.018 \\
& F$_1$-score (Negative) & 0.96 $\pm$ 0.004 & 0.96 $\pm$ 0.005 & 0.96 $\pm$ 0.004 & 0.97 $\pm$ 0.003 & 0.94 $\pm$ 0.006\\
& HSS$_1$ &0.92 $\pm$ 0.008 & 0.92 $\pm$ 0.009 & 0.91 $\pm$ 0.008 & 0.95 $\pm$ 0.007 & 0.88 $\pm$ 0.012\\
& HSS$_2$ &0.81 $\pm$ 0.019 & 0.80 $\pm$ 0.021 & 0.80 $\pm$ 0.018 &0.85 $\pm$ 0.017 & 0.75 $\pm$ 0.024 \\
& GS & 0.69 $\pm$ 0.027 &  0.67 $\pm$ 0.030 & 0.67 $\pm$ 0.025 &0.74 $\pm$ 0.026 & 0.61 $\pm$ 0.031 \\
& TSS &0.84 $\pm$ 0.022 & 0.82 $\pm$ 0.024 & 0.82 $\pm$ 0.021 & 0.88 $\pm$ 0.020 & 0.77 $\pm$ 0.027\\
\hline
\end{tabular}
  \end{center}
\end{table*}

\begin{table*}[t]
\caption[]{Flare prediction capabilities obtained from our complex classifier models namely support vector machine and multilayer perceptron for five different data sets. First three data sets namely loop24span12, loop24span24, loop24span0 correspond to the data sets with same forecasting window of 24 hour but different span time of 12 hour, 48 hour and zero hour respectively. While last two data sets namely loop12span12 and loop48span12 correspond to the data sets with same span time of 12 hour but different forecasting window of 12 hour and 48 hour respectively. We consider all magnetic parameters to create the data set. }
 \begin{center}\begin{tabular}{c|c|c|c|c|c|c|}

   &  & loop24span12 & loop24span24& loop24span0 & loop12span12& loop48span12 \\
  \hline
Support   & Accuracy & 0.96 $\pm$ 0.019 & 0.97 $\pm$ 0.019 &0.97 $\pm$ 0.005 & 0.97 $\pm$ 0.005 & 0.94 $\pm$ 0.008\\
Vector  & Precision (Positive) & 0.90 $\pm$ 0.019 &  0.91 $\pm$ 0.019 & 0.88 $\pm$ 0.019 & 0.89 $\pm$ 0.022 & 0.84 $\pm$ 0.023 \\
Machine  & Precision (Negative) &0.98 $\pm$ 0.005 & 0.98 $\pm$ 0.005 & 0.98 $\pm$ 0.004 & 0.98 $\pm$ 0.003 & 0.97 $\pm$ 0.006\\
(SVM)  & Recall (Positive) & 0.92 $\pm$ 0.021& 0.92 $\pm$ 0.022 & 0.95 $\pm$ 0.014 & 0.92 $\pm$ 0.013 & 0.90 $\pm$ 0.021\\
& Recall (Negative) &0.97 $\pm$ 0.005 & 0.97 $\pm$ 0.005 & 0.97 $\pm$ 0.006 & 0.98 $\pm$ 0.003 & 0.95 $\pm$ 0.009 \\
& F$_1$-score (Positive) &0.91 $\pm$ 0.015 &0.92 $\pm$ 0.014 &0.92 $\pm$ 0.012 & 0.92 $\pm$ 0.013 & 0.87 $\pm$ 0.016 \\
& F$_1$-score (Negative) & 0.98 $\pm$ 0.004 & 0.98 $\pm$ 0.004 & 0.98 $\pm$ 0.003 & 0.98 $\pm$ 0.003 & 0.96 $\pm$ 0.005\\
& HSS$_1$ &0.95 $\pm$ 0.008 & 0.96 $\pm$ 0.007 & 0.96 $\pm$ 0.007 & 0.96 $\pm$ 0.006 & 0.92 $\pm$ 0.010\\
& HSS$_2$ &0.89 $\pm$ 0.019 & 0.89 $\pm$ 0.018 & 0.90 $\pm$ 0.015 &0.90 $\pm$ 0.016 & 0.82 $\pm$ 0.021 \\
& GS & 0.80 $\pm$ 0.030 &  0.81 $\pm$ 0.029 & 0.81 $\pm$ 0.025 &0.82 $\pm$ 0.026 & 0.71 $\pm$ 0.030 \\
& TSS &0.90 $\pm$ 0.022 & 0.90 $\pm$ 0.022 & 0.92 $\pm$ 0.014 & 0.93 $\pm$ 0.015 & 0.85 $\pm$ 0.021\\
\hline
Multilayer   & Accuracy & 0.96 $\pm$ 0.017 & 0.95 $\pm$ 0.029 &0.96 $\pm$ 0.023 & 0.97 $\pm$ 0.009 & 0.92 $\pm$ 0.046\\
Perceptron  & Precision (Positive) & 0.89 $\pm$ 0.071 &  0.88 $\pm$ 0.107 & 0.89 $\pm$ 0.075 & 0.92 $\pm$ 0.046 & 0.81 $\pm$ 0.105 \\
(MLP)  & Precision (Negative) &0.98 $\pm$ 0.005 & 0.98 $\pm$ 0.009 & 0.98 $\pm$ 0.007 & 0.98 $\pm$ 0.003 & 0.96 $\pm$ 0.017\\
  & Recall (Positive) & 0.93 $\pm$ 0.040& 0.93 $\pm$ 0.056 & 0.94 $\pm$ 0.030 & 0.95 $\pm$ 0.015 & 0.87 $\pm$ 0.064\\
& Recall (Negative) &0.96 $\pm$ 0.027 & 0.96 $\pm$ 0.043 & 0.97 $\pm$ 0.031 & 0.98 $\pm$ 0.013 & 0.93 $\pm$ 0.069 \\
& F$_1$-score (Positive) &0.90 $\pm$ 0.032 &0.89 $\pm$ 0.052 &0.91 $\pm$ 0.039 & 0.93 $\pm$ 0.021 & 0.84 $\pm$ 0.058 \\
& F$_1$-score (Negative) & 0.97 $\pm$ 0.012 & 0.97 $\pm$ 0.020 & 0.97 $\pm$ 0.016 & 0.98 $\pm$ 0.006 & 0.94 $\pm$ 0.037\\
& HSS$_1$ &0.95 $\pm$ 0.021 & 0.94 $\pm$ 0.036 & 0.95 $\pm$ 0.027 & 0.97 $\pm$ 0.011 & 0.89 $\pm$ 0.060\\
& HSS$_2$ &0.87 $\pm$ 0.041 & 0.87 $\pm$ 0.070 & 0.89 $\pm$ 0.053 &0.92 $\pm$ 0.027 & 0.78 $\pm$ 0.077 \\
& GS & 0.78 $\pm$ 0.065 &  0.77 $\pm$ 0.101 & 0.81 $\pm$ 0.064 &0.84 $\pm$ 0.045 & 0.65 $\pm$ 0.101 \\
& TSS &0.90 $\pm$ 0.030 & 0.89 $\pm$ 0.041 & 0.91 $\pm$ 0.028 & 0.93 $\pm$ 0.012 & 0.81 $\pm$ 0.057\\
\hline
\end{tabular}
  \end{center}
\end{table*}

  \begin{table*}[t]
\caption[]{Flare prediction capabilities obtained from different classifiers for loop24span12 data set}
 \begin{center}\begin{tabular}{c|c|c|c|c|c|c|}
 \hline
 \multicolumn{7}{c} {Performance by other classifiers on our loop24span12 dataset}\\
 \hline
   Our Classifiers:  & LR & SVM& MLP & KNN& Random Forest & Naive Bayes\\
  \hline
  Accuracy & 0.95 $\pm$ 0.023 & 0.96 $\pm$ 0.019 &0.96 $\pm$ 0.017 & 0.95 $\pm$ 0.006 & 0.96 $\pm$ 0.006 & 0.94 $\pm$ 0.008 \\
Precision (Positive) & 0.84 $\pm$ 0.023 &  0.90 $\pm$ 0.019 & 0.89 $\pm$ 0.071 & 0.90 $\pm$ 0.018 & 0.93 $\pm$ 0.019 & 0.82 $\pm$ 0.033 \\
Precision (Negative) &0.98 $\pm$ 0.005 & 0.98 $\pm$ 0.005 & 0.98 $\pm$ 0.010 & 0.96 $\pm$ 0.006 & 0.97 $\pm$ 0.006 & 0.97 $\pm$ 0.006\\
Recall (Positive) & 0.93 $\pm$ 0.018& 0.92 $\pm$ 0.021 & 0.93 $\pm$ 0.040 & 0.86 $\pm$ 0.025 & 0.87 $\pm$ 0.024 & 0.88 $\pm$ 0.023\\
Recall (Negative) &0.95 $\pm$ 0.008 & 0.97 $\pm$ 0.005 & 0.96 $\pm$ 0.027 & 0.98 $\pm$ 0.005 & 0.98 $\pm$ 0.005 & 0.96 $\pm$ 0.005 \\
 F$_1$-score (Positive) &0.88 $\pm$ 0.014 &0.91 $\pm$ 0.015 &0.90 $\pm$ 0.032 & 0.85 $\pm$ 0.017 & 0.90 $\pm$ 0.015 & 0.85 $\pm$ 0.017 \\
 F$_1$-score (Negative) & 0.97 $\pm$ 0.004 & 0.98 $\pm$ 0.004 & 0.97 $\pm$ 0.012 & 0.97 $\pm$ 0.004 & 0.98 $\pm$ 0.005 & 0.96 $\pm$ 0.005\\
 HSS$_1$ &0.93 $\pm$ 0.008 & 0.96 $\pm$ 0.008 & 0.95 $\pm$ 0.022 & 0.94 $\pm$ 0.007 & 0.95 $\pm$ 0.007 & 0.92 $\pm$ 0.010\\
 HSS$_2$ &0.84 $\pm$ 0.018 & 0.89 $\pm$ 0.019 & 0.87 $\pm$ 0.044 &0.85 $\pm$ 0.019 & 0.87 $\pm$ 0.019 & 0.81 $\pm$ 0.022 \\
 GS & 0.73 $\pm$ 0.027 &  0.80 $\pm$ 0.030 & 0.78 $\pm$ 0.065 &0.74 $\pm$ 0.029 & 0.78 $\pm$ 0.029 & 0.68 $\pm$ 0.031 \\
 TSS &0.88 $\pm$ 0.017 & 0.90 $\pm$ 0.022 & 0.90 $\pm$ 0.030 & 0.83 $\pm$ 0.025 & 0.86 $\pm$ 0.024 & 0.84 $\pm$ 0.020\\
\hline
\end{tabular}
  \end{center}
\end{table*}

\begin{figure*}
        \centering
        \includegraphics[width=0.8\textwidth]{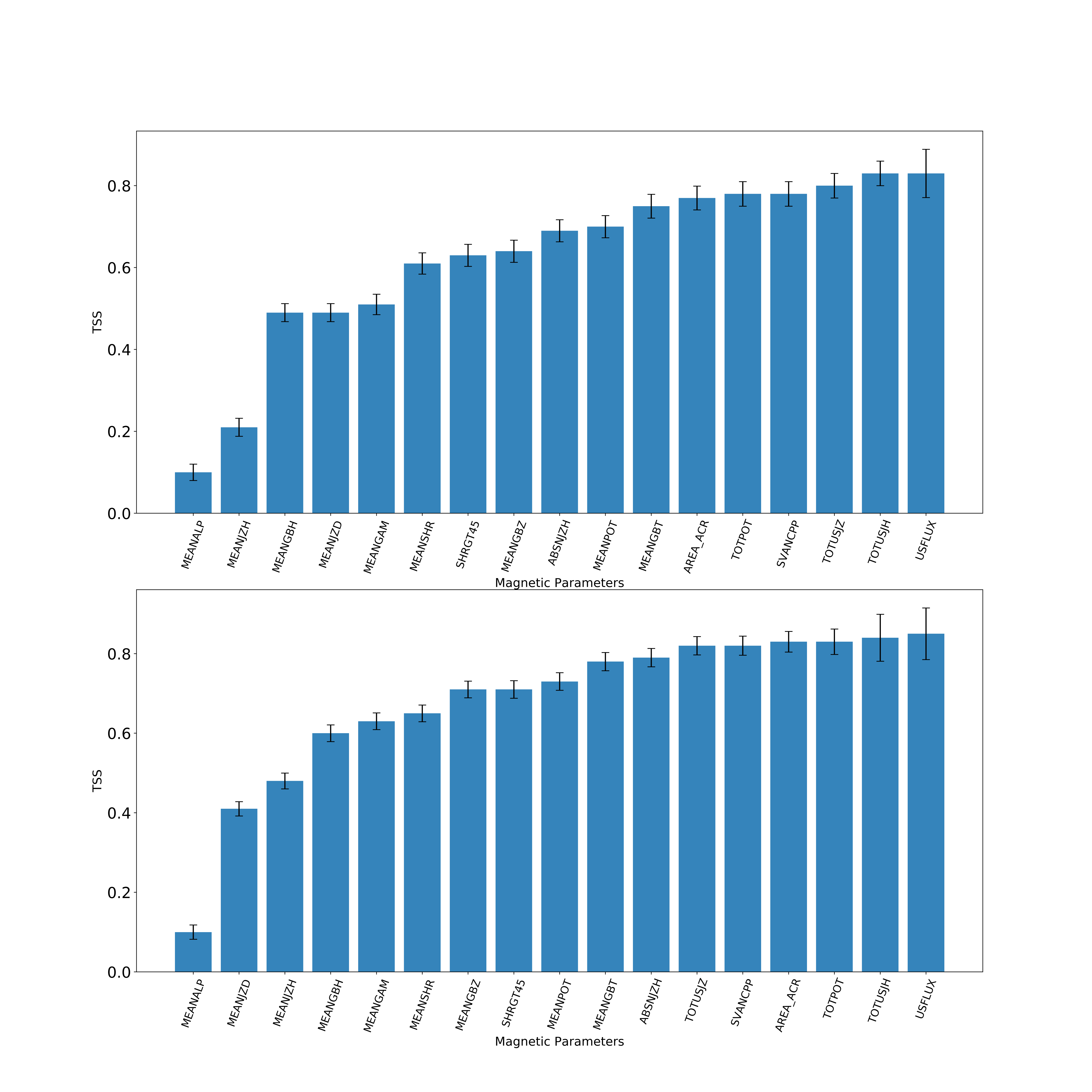}
        \caption{Top bar plot shows the distribution of TSS values after running the logistic regression (baseline) classifier on the summarized time series of individual active region parameters. Bottom bar plot represents the same but used support vector machine algorithm. This plots clearly show that top few parameters, specifically, total unsigned magnetic flux and the total unsigned magnetic helicity have the best distinguishing capability} 
        \label{fig:2}
\end{figure*}
 
 \begin{figure*}
        \centering
        \includegraphics[width=0.8\textwidth]{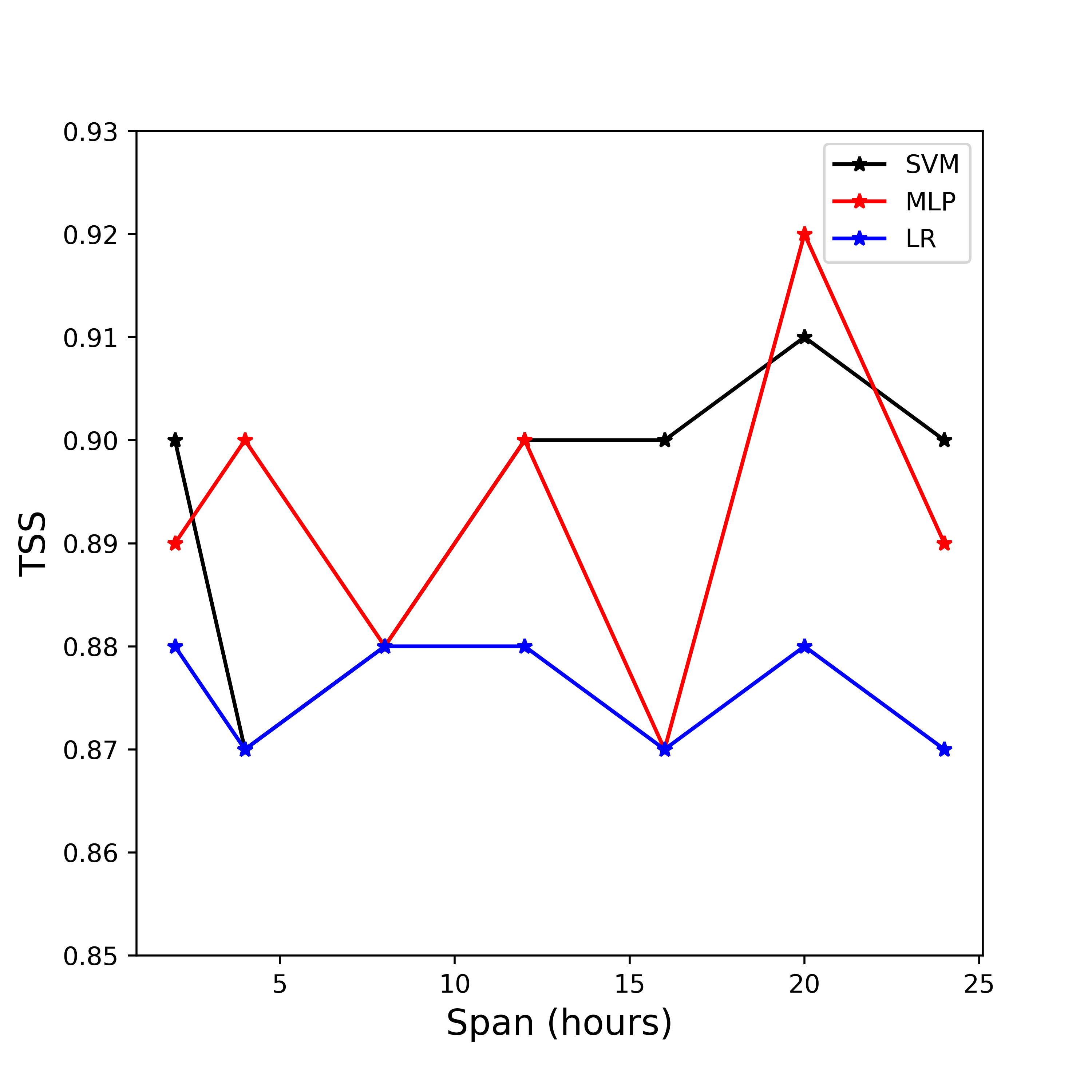}
        \caption{Variation of the true skill statistic (TSS) after running the support vector machine and multilayer perceptron over the data sets with different span but fixed 24 hour forecasting window, i.e. loopback.}
        \label{fig:2}
\end{figure*}

\subsubsection{Results}

We use the active region vector magnetogram data from June 2010 to December 2017 for this study. After generating the dataset, we perform some preprocessing in our data set due to the presence of some missing values. We simply replace them with the mean of the corresponding features. It is also necessary to normalize the dataset before training as it will transform the ranges of all feature values into a uniform range. In this study, we use the zero-one data transformation technique for normalization. We randomly divide our dataset into a training set ($70 \%$) and a testing data set ($30 \%$). We maintain the same class ratio (N/P) in both training and testing data set following the prescription of \cite{bobr15} and \cite{ahme13}. Please note that we do not include C class flares in our positive data set. Our data set is highly imbalanced as flaring regions are much less compared to non-flaring ones. Oversampling and undersampling are the two well-known strategies to make an imbalanced data set balanced. While oversampling involves the strategy of adding more positive examples in the dataset, undersampling involves the strategy of removing the majority negative examples from the dataset. However, both methods have some limitations. In oversampling, the addition of many replicas of positive examples may cause the model to memorize the pattern and makes the model prone to overfitting. On the other hand, we are removing many negative examples in the undersampling strategy, thus the computer is not learning from the entire data set.

In this study, we use the concept of weighted techniques to tackle the issue of class imbalance. In the weighted techniques, we provide more weights to the classes which we aim to predict. \cite{azim19} found that weighted techniques work better to tackle the class imbalance issue compared to oversampling and undersampling techniques in case of flare prediction studies. The weighted technique is also free from the limitations of undersampling and oversampling techniques. In our study, we aim to predict flaring active regions which are the minority one. We have provided more weight to the flaring active region compared to the non-flaring one such that our classifier does not focus exclusively on the non-flaring class which is the majority one. In our SVM classifier, We set the cost parameter of the flaring class higher compared to the non-flaring one to tackle the issue of class imbalance. We also use a similar technique for our baseline classifier to address the issue of class imbalance. In the case of our other classifier multilayer perceptron, we use the weight balancing technique to encounter the issue of class imbalance. In the weight balancing technique, we alter the weights of each training data during the computation of loss function. Generally, both positive and negative classes carry the same weight 1.0 in the computation of loss function. As our primary aim here is to predict minority classes (flaring one), we provide more weight to the flaring classes in the calculation of loss function compared to the non-flaring one. 

Table 1 lists the performance metrics found after running the baseline weighted logistic regression classifier on the five different data sets. We have provided the means and standard deviations in the table by repeating training and testing phases few times. We use a 24-hour forecasting window for the data sets namely loop24span12, loop24span24, and loop24span0. We use 12 and 24 hours time evolution of magnetic parameters respectively for the data sets loop24span12 and loop24span24 as training purposes (span 12 hours and 24 hours). While we use no time evolution of magnetic parameters for the data set loop24span0 as training purposes. The loop24span0 data set is almost similar like \cite{bobr15} but with larger time duration (June 2010 to December 2017). We use 12 hours time evolution of magnetic parameters respectively for the data sets namely loop12span12 and loop48span12 as training purpose but we use different forecasting windows. The forecasting window is 12 hours for the dataset loop12span12; while the forecasting window for the dataset loop48span12 is 48 hours.
 We found a very good TSS value for all five cases (see Table 1). We get the maximum TSS value (~0.92) for the data set loop12span12. Interestingly we also find very good TSS value for the data set loop24span0 where we do not use any time evolution for training. In summary, time evolution does not have much impact on the performance of a classifier. We also find a good TSS value for the data set loop48span12. Our baseline model result indicates that it is possible to distinguish a flaring active region from the non-flaring one. 

Next, we use two more advanced classifiers namely support vector machine and multilayer perceptron for our experiment. We wanted to know whether our selected advanced classifiers can improve the performance significantly compared to our baseline model. These two advanced classifiers can also separate non-linearly separable classes. We tune the hyperparameters of the support vector machine (SVM) algorithm by using a grid search algorithm provided by the Scikit-Learn software package \citep{pedr12}. This algorithm finds the best combination of hyperparameters after performing an exhaustive search over a predefined set of hyperparameters. Our SVM hyperparameters namely, regularization parameter (C) varies between 0.001 and 10 and the RBF parameter $\gamma$ varies between 0.001 and 1. We use {\it binary cross-entropy} as a loss function and {\it rmsprop} as an optimizer for our multilayer perceptron algorithm. We also use {\it Keras} {\it l2} regularizer as a kernel regularizer for our MLP algorithm. Table 2 lists the eleven performance metrics found after running two different classifiers, namely support vector machine and multilayer perceptron, on the five different data sets. We find that both the support vector machine and the multilayer perceptron algorithm performs well to distinguish the flaring active regions from the non-flaring one. We also note that TSS values obtained from SVM and MLP classifiers are higher compared to TSS values obtained from our baseline logistic regression classifier. In summary, advanced classifiers perform well compared to the baseline model. However, the performance of the advanced classifiers is very close to the performance of the baseline model. The baseline logistic regression classifier (linear classifier) works very well in distinguishing flaring and non-flaring regions.

One may pose the question-- which magnetic parameter is most critical for distinguishing flaring and non-flaring active regions? We follow the suggestions of \cite{hamd17} to find out the best active region parameter. In this study, we use the time evolution of magnetic parameters for training machine learning algorithms. The magnetic parameter whose corresponding time evolution gives maximum true skill statistic (TSS) after the classification by logistic regression (baseline) and SVM algorithm is considered to be the best among all active region magnetic parameters in terms of distinguishing capability. We use the data set loop24span12 for this experiment. For this purpose, we train both SVM and LR algorithm by using the time evolution of a single magnetic parameter at a time and measure the TSS value for each case. The bar plot of Figure 5 shows that both the total unsigned magnetic helicity and total unsigned magnetic flux achieve maximum TSS. These results indicate that the time evolution of the total unsigned magnetic helicity and the total unsigned magnetic flux is the best indicator in terms of distinguishing capability. \cite{bobr15} also found that the total unsigned magnetic helicity parameter is the best active region parameter based on the Fisher criterion. However, we find that the time evolution of total unsigned magnetic flux also has an equally good distinguishing capability. We also note that time evolution of some other parameters namely total photospheric magnetic free energy density, total unsigned vertical current, and AREA$\_$ACR also have a good distinguishing capability (see Fig. $5$). Time evolution of total unsigned magnetic flux is an indicator of flux cancellation and emergence on the solar surface. On the other hand, the time evolution of magnetic helicity indicates the magnetic complexity of the active region. Our finding in terms of critical active region parameter is consistent with the earlier theoretical and observational findings.

Previous studies suggested that there will be no overfitting if we use twelve to eighteen magnetic parameters \citep{bobr16,ince18}. Because of this, we did not apply any feature selection criterion before the application of a machine learning algorithm. Now, we select only the top five best magnetic parameters for our study. Our selected top five best magnetic parameters are total unsigned magnetic flux (USFLUX), total unsigned current helicity (TOTUSJH), total unsigned vertical current (TOTUSJZ), total photospheric magnetic free energy density (TOTPOT), and the area of strong field pixels in the active region (AREA$\_$ACR). We train our baseline logistic regression algorithm by using the time evolution of these five magnetic parameters. The bottom part of Table $1$ lists the performance metrics found after running the classifiers over the data sets which considers only the best five magnetic parameters. We find that our baseline classifier performs quite well in terms of distinguishing capability even if we consider only the top five magnetic parameters. We also note that TSS values obtained by using only the top five magnetic parameters are very close to the values obtained by considering all magnetic parameters.

How does the classifier performance change with the forecast window? To investigate this issue, we create the data sets with different forecasting windows (loop) but the same span. Table 1 shows the classification metrics obtained after running the classifier over three data sets, namely, loop12span12, loop24span12, and loop48span12. So, we have three data sets namely loop12span12, loop24span12, and loop48span12 for this study. We use a 12-hour time evolution of magnetic parameters for training in all cases. Table 1 shows that there is a decreasing trend of TSS value with our selected forecasting windows.  However, we get very good TSS values for all three data sets. Our selected forecasting windows have very little impact on the performance of the classifier.

How does the time evolution of magnetic parameters for different time windows (span) change the performance of the classifier? For this purpose, we consider data sets having a fixed forecasting window of 24 hours but different span windows. Figure 6 shows the change of TSS values after running the SVM and MLP classifiers over these data sets. We do not find any increasing or decreasing trend of TSS values with the span size. Span size may not have much impact on the performance of the classifier.

Table 3 shows the different performance metrics found after running some other classifiers namely, random forest, Knn, Naive Bayes on the loop24span12 data set. We have also listed the results of our previous three classifiers for comparison. The hyperparameters for our other classifiers are the following: Random Forest (n estimators=10, max depth=None, criterion=gini, class weight=balanced), Naive Bayes (Priors=None), KNN (number of neighbors k=1, distance= Eucledian). A comparison between the results obtained from different classifiers confirms the robustness of our model in distinguishing flaring and non-flaring active regions. We also note that our baseline classifier Logistic Regression (which is easily interpretable) have very good performance. The code for this study can be found in our github repository (\href{https://github.com/soumitrahazra/Flaring_Region_Prediction}{\textcolor{blue}{http://gitub.com/soumitrahazra/Flaring\_Region\_Prediction}}).

 \section{Summary} 
We have performed a comparison study between eruptive and non-eruptive active regions in terms of the time evolution of photospheric magnetic parameters. We first performed this study manually. We have selected two eruptive active regions and one non-eruptive active region to find out the difference between the time evolution of magnetic complexity in eruptive and non-eruptive active regions. We find a significant difference between the eruptive and non-eruptive active region in terms of both strength and time evolution of photospheric magnetic parameters. All of our selected magnetic parameters, namely total unsigned magnetic flux, total unsigned helicity, total photospheric magnetic free energy density, and the total unsigned vertical current have a much higher value in case of eruptive active regions compared to the non-eruptive active region. We find the signature of flux emergence and cancellation in the time evolution of the total unsigned magnetic flux. Time evolution of the total unsigned current helicity shows a strong indication of both helicity accumulation and annihilation at the onset of a solar flare.

As it is not possible to analyze all flaring events manually due to a large amount of solar data, we have used machine learning algorithms to distinguish eruptive active regions from the non-eruptive one. We use the time evolution of photospheric magnetic parameters to train our baseline machine learning algorithm logistic regression. Motivated by \cite{bloom12} and \cite{bobr15}, we have selected true skill statistic (TSS) as a performance measure of our forecasting algorithm as TSS does not depend on the class imbalance. Solar flare prediction is a highly imbalanced problem as there is a fewer flaring active region compared to non-flaring ones. We obtain high TSS values for our baseline algorithm. A higher value of TSS also implies a lower false-negative rate which is very important for the forecasting study. We have then compared our baseline result with the results obtained from more advanced classifiers, namely, support vector machine and multilayer perceptron. We find that the use of advanced machine learning algorithms improves performance. But the improvement is not much significant.

We note that our TSS value is higher compared to previous studies. This may be due to our data set (as the data set is different from previous studies) or may be due to our choice of training the computer by the evolution of photospheric magnetic parameters. \cite{mart18} describes the importance of the "benchmark data set" for this kind of prediction studies. Preparation of the "benchmark data set" involves many processes like gathering a large amount of data, cleaning up the data and balancing these data. We have also performed one study where we did not use the time evolution of photospheric magnetic parameters to train the machine learning algorithm. We found a very good TSS value in that case. That's why we are not sure whether time evolution of magnetic parameters is really important for flare forecasting study. However, we note that we describe the time evolution by eight statistical parameters -- which may not be a very good way to describe the time evolution (but this is the simplest one). We have not considered any statistical quantities to represent seasonal or periodic features in this study. Features generated from fast Fourier transform (FFT) of the time series will be helpful to capture the information of periodic features of the time series. Different characteristics of the time series carry some important pieces of information. There is also no established work at this moment which clearly determines the effective statistical features for improved flare forecasting. We leave this investigation for our future studies.

Next, we try to find out the common critical magnetic parameter which will clearly distinguish the eruptive active regions from the non-eruptive one. We find that time evolution of total unsigned magnetic helicity and total unsigned magnetic flux have a very high distinguishing capability. We note that the time evolution of unsigned magnetic flux is an indicator of flux emergence and cancellation on the solar surface; while time evolution of unsigned magnetic helicity represents the helicity accumulation and cancellation. Previous theoretical models describe the importance of both this mechanism in detail at the onset phase of the solar model. Our manual analysis with three active regions also supports this view. Now, we want to compare this result with some earlier studies. \cite{bobr15} found that total unsigned magnetic flux has the best-distinguishing capability compared to others. However, they did not find any significant distinguishing capability of the total unsigned magnetic flux. Please note that they did not use the time evolution of magnetic parameters to train the machine learning algorithm. \cite{ma17} used the univariate time series clustering and multivariate time series decision tree for the purpose of flare prediction and found that both the total unsigned magnetic flux and the total unsigned magnetic helicity have very high distinguishing capability compared to other.

We also find very high TSS value when we consider only the evolution of the top five magnetic parameters for the training of machine learning algorithms. In summary, although we are not able to find a single critical magnetic parameter, we find that a combination of the top few magnetic parameters will give us almost similar distinguishing capability. This result is consistent with earlier studies. Earlier studies also found that the top few magnetic parameters can produce the forecasting capability comparable to their entire data set \citep{leka03, leka07, ahme13, bobr15, hamd17}. We note that "total" parameters are more valuable in flare study compared to the "mean" one. \cite{wels09} also found that extensive magnetic parameters (whose value increases with size) have a stronger correlation with the flare productivity compared to the intensive one (value does not increase with size). This result indicates that larger and complex active regions are more flare prone compared to a smaller one.
  \begin{acknowledgements}
 The Center of Excellence in Space Sciences India (CESSI) is supported by the Ministry of Human Resource Development, Government of India. We also thank Georgia State University data mining group for the discussion and suggestions.
\end{acknowledgements}

%-------------------------------------------------------------------

%%%%%%%%%%%%%%%%%%%%%%%%%%%%%%%%%%%%%%%%%%%%%%%%%%



%\bibliographystyle{aa}
%\bibliography{reference} % if your bibtex file is called example.bib

\begin{thebibliography}{66}
\expandafter\ifx\csname natexlab\endcsname\relax\def\natexlab#1{#1}\fi

\bibitem[{{Abramenko} {et~al.}(1996){Abramenko}, {Wang}, \&
  {Yurchishin}}]{abra96}
{Abramenko}, V.~I., {Wang}, T., \& {Yurchishin}, V.~B. 1996, \solphys, 168, 75

\bibitem[{{Aggarwal} {et~al.}(2018){Aggarwal}, {Schanche}, {Reeves}, {Kempton},
  \& {Angryk}}]{agga18}
{Aggarwal}, A., {Schanche}, N., {Reeves}, K.~K., {Kempton}, D., \& {Angryk}, R.
  2018, \apjs, 236, 15

\bibitem[{{Ahmadzadeh} {et~al.}(2019){Ahmadzadeh}, {Hostetter}, {Aydin},
  {Georgoulis}, {Kempton}, {Mahajan}, \& {Angryk}}]{azim19}
{Ahmadzadeh}, A., {Hostetter}, M., {Aydin}, B., {et~al.} 2019, arXiv e-prints,
  arXiv:1911.09061

\bibitem[{{Ahmed} {et~al.}(2013){Ahmed}, {Qahwaji}, {Colak}, {Higgins},
  {Gallagher}, \& {Bloomfield}}]{ahme13}
{Ahmed}, O.~W., {Qahwaji}, R., {Colak}, T., {et~al.} 2013, \solphys, 283, 157

\bibitem[{{Amari} {et~al.}(2010){Amari}, {Aly}, {Mikic}, \& {Linker}}]{amar10}
{Amari}, T., {Aly}, J.~J., {Mikic}, Z., \& {Linker}, J. 2010, \apjl, 717, L26

\bibitem[{{Ambastha} {et~al.}(1993){Ambastha}, {Hagyard}, \& {West}}]{amba93}
{Ambastha}, A., {Hagyard}, M.~J., \& {West}, E.~A. 1993, \solphys, 148, 277

\bibitem[{{Barnes} \& {Leka}(2008)}]{barn08}
{Barnes}, G. \& {Leka}, K.~D. 2008, \apjl, 688, L107

\bibitem[{{Berger} \& {Field}(1984)}]{berg84}
{Berger}, M.~A. \& {Field}, G.~B. 1984, Journal of Fluid Mechanics, 147, 133

\bibitem[{{Bloomfield} {et~al.}(2012){Bloomfield}, {Higgins}, {McAteer}, \&
  {Gallagher}}]{bloom12}
{Bloomfield}, D.~S., {Higgins}, P.~A., {McAteer}, R.~T.~J., \& {Gallagher},
  P.~T. 2012, \apjl, 747, L41

\bibitem[{{Bobra} \& {Couvidat}(2015)}]{bobr15}
{Bobra}, M.~G. \& {Couvidat}, S. 2015, \apj, 798, 135

\bibitem[{{Bobra} \& {Ilonidis}(2016)}]{bobr16}
{Bobra}, M.~G. \& {Ilonidis}, S. 2016, \apj, 821, 127

\bibitem[{{Bobra} {et~al.}(2014){Bobra}, {Sun}, {Hoeksema}, {Turmon}, {Liu},
  {Hayashi}, {Barnes}, \& {Leka}}]{bobr14}
{Bobra}, M.~G., {Sun}, X., {Hoeksema}, J.~T., {et~al.} 2014, \solphys, 289,
  3549

\bibitem[{{Burtseva} \& {Petrie}(2013)}]{burt13}
{Burtseva}, O. \& {Petrie}, G. 2013, \solphys, 283, 429

\bibitem[{{Crouch} {et~al.}(2009){Crouch}, {Barnes}, \& {Leka}}]{crou09}
{Crouch}, A.~D., {Barnes}, G., \& {Leka}, K.~D. 2009, \solphys, 260, 271

\bibitem[{{Dhuri} {et~al.}(2019){Dhuri}, {Hanasoge}, \& {Cheung}}]{dhur19}
{Dhuri}, D.~B., {Hanasoge}, S.~M., \& {Cheung}, M. C.~M. 2019, Proceedings of
  the National Academy of Science, 116, 11141

\bibitem[{{Filali Boubrahimi} \& {Angryk}(2018)}]{boub18}
{Filali Boubrahimi}, S. \& {Angryk}, R. 2018, in 2018 IEEE First International
  Conference on Artificial Intelligence and Knowledge Engineering (AIKE),
  162--163

\bibitem[{{Florios} {et~al.}(2018){Florios}, {Kontogiannis}, {Park}, {Guerra},
  {Benvenuto}, {Bloomfield}, \& {Georgoulis}}]{flor18}
{Florios}, K., {Kontogiannis}, I., {Park}, S.-H., {et~al.} 2018, \solphys, 293,
  28

\bibitem[{{Gaizauskas} {et~al.}(1997){Gaizauskas}, {Zirker}, {Sweetland}, \&
  {Kovacs}}]{gaiz97}
{Gaizauskas}, V., {Zirker}, J.~B., {Sweetland}, C., \& {Kovacs}, A. 1997, \apj,
  479, 448

\bibitem[{{Hamdi} {et~al.}(2017){Hamdi}, {Kempton}, {Ma}, {Boubrahimi}, \&
  {Angryk}}]{hamd17}
{Hamdi}, S.~M., {Kempton}, D., {Ma}, R., {Boubrahimi}, S.~F., \& {Angryk},
  R.~A. 2017, in 2017 IEEE International Conference on Big Data (Big Data),
  2543--2551

\bibitem[{{Hazra} {et~al.}(2018){Hazra}, {Mahajan}, {Douglas}, \&
  {Martens}}]{hazr18}
{Hazra}, S., {Mahajan}, S.~S., {Douglas}, William~Keith, J., \& {Martens}, P.
  C.~H. 2018, \apj, 865, 108

\bibitem[{{Hazra} {et~al.}(2015){Hazra}, {Nandy}, \& {Ravindra}}]{hazr15}
{Hazra}, S., {Nandy}, D., \& {Ravindra}, B. 2015, \solphys, 290, 771

\bibitem[{{Holder} {et~al.}(2004){Holder}, {Canfield}, {McMullen}, {Nandy},
  {Howard}, \& {Pevtsov}}]{hold04}
{Holder}, Z.~A., {Canfield}, R.~C., {McMullen}, R.~A., {et~al.} 2004, \apj,
  611, 1149

\bibitem[{{Inceoglu} {et~al.}(2018){Inceoglu}, {Jeppesen}, {Kongstad},
  {Hern{\'a}ndez Marcano}, {Jacobsen}, \& {Karoff}}]{ince18}
{Inceoglu}, F., {Jeppesen}, J.~H., {Kongstad}, P., {et~al.} 2018, \apj, 861,
  128

\bibitem[{{Kusano} {et~al.}(2002){Kusano}, {Maeshiro}, {Yokoyama}, \&
  {Sakurai}}]{kusa02}
{Kusano}, K., {Maeshiro}, T., {Yokoyama}, T., \& {Sakurai}, T. 2002, \apj, 577,
  501

\bibitem[{{Kusano} {et~al.}(2004){Kusano}, {Maeshiro}, {Yokoyama}, \&
  {Sakurai}}]{kusa04}
{Kusano}, K., {Maeshiro}, T., {Yokoyama}, T., \& {Sakurai}, T. 2004, \apj, 610,
  537

\bibitem[{{Kusano} {et~al.}(1995){Kusano}, {Suzuki}, \& {Nishikawa}}]{kusa95}
{Kusano}, K., {Suzuki}, Y., \& {Nishikawa}, K. 1995, \apj, 441, 942

\bibitem[{{Kusano} {et~al.}(2003){Kusano}, {Yokoyama}, {Maeshiro}, \&
  {Sakurai}}]{kusa03}
{Kusano}, K., {Yokoyama}, T., {Maeshiro}, T., \& {Sakurai}, T. 2003, Advances
  in Space Research, 32, 1931

\bibitem[{{Leka} \& {Barnes}(2003)}]{leka03}
{Leka}, K.~D. \& {Barnes}, G. 2003, \apj, 595, 1296

\bibitem[{{Leka} \& {Barnes}(2007)}]{leka07}
{Leka}, K.~D. \& {Barnes}, G. 2007, \apj, 656, 1173

\bibitem[{{Leka} {et~al.}(1993){Leka}, {Canfield}, {McClymont}, {de La
  Beaujardiere}, {Fan}, \& {Tang}}]{leka93}
{Leka}, K.~D., {Canfield}, R.~C., {McClymont}, A.~N., {et~al.} 1993, \apj, 411,
  370

\bibitem[{{Livi} {et~al.}(1989){Livi}, {Martin}, {Wang}, \& {Ai}}]{livi89}
{Livi}, S. H.~B., {Martin}, S., {Wang}, H., \& {Ai}, G. 1989, \solphys, 121,
  197

\bibitem[{{Low}(1977)}]{low77}
{Low}, B.~C. 1977, \apj, 217, 988

\bibitem[{{Ma} {et~al.}(2017){Ma}, {Boubrahimi}, {Hamdi}, \& {Angryk}}]{ma17}
{Ma}, R., {Boubrahimi}, S.~F., {Hamdi}, S.~M., \& {Angryk}, R.~A. 2017, in 2017
  IEEE International Conference on Big Data (Big Data), 2569--2578

\bibitem[{{Martens} \& {Angryk}(2018)}]{mart18}
{Martens}, P.~C. \& {Angryk}, R.~A. 2018, in IAU Symposium, Vol. 335, Space
  Weather of the Heliosphere: Processes and Forecasts, ed. C.~{Foullon} \&
  O.~E. {Malandraki}, 344--347

\bibitem[{{Martens} \& {Zwaan}(2001)}]{mart01}
{Martens}, P.~C. \& {Zwaan}, C. 2001, \apj, 558, 872

\bibitem[{{Martin} {et~al.}(1994){Martin}, {Bilimoria}, \&
  {Tracadas}}]{marti94}
{Martin}, S.~F., {Bilimoria}, R., \& {Tracadas}, P.~W. 1994, in NATO Advanced
  Science Institutes (ASI) Series C, Vol. 433, NATO Advanced Science Institutes
  (ASI) Series C, ed. R.~J. {Rutten} \& C.~J. {Schrijver}, 303

\bibitem[{{Martin} {et~al.}(1985){Martin}, {Livi}, \& {Wang}}]{marti85}
{Martin}, S.~F., {Livi}, S.~H.~B., \& {Wang}, J. 1985, Max Planck Institut fur
  Astrophysik Report, 212, 179

\bibitem[{{Mason} \& {Hoeksema}(2010)}]{maso10}
{Mason}, J.~P. \& {Hoeksema}, J.~T. 2010, \apj, 723, 634

\bibitem[{{Metcalf}(1994)}]{metc94}
{Metcalf}, T.~R. 1994, \solphys, 155, 235

\bibitem[{{Metcalf} {et~al.}(2005){Metcalf}, {Leka}, \& {Mickey}}]{metc05}
{Metcalf}, T.~R., {Leka}, K.~D., \& {Mickey}, D.~L. 2005, \apjl, 623, L53

\bibitem[{{Moon} {et~al.}(2002){Moon}, {Chae}, {Choe}, {Wang}, {Park}, {Yun},
  {Yurchyshyn}, \& {Goode}}]{moon02}
{Moon}, Y.~J., {Chae}, J., {Choe}, G.~S., {et~al.} 2002, \apj, 574, 1066

\bibitem[{{Nishizuka} {et~al.}(2017){Nishizuka}, {Sugiura}, {Kubo}, {Den},
  {Watari}, \& {Ishii}}]{nish17}
{Nishizuka}, N., {Sugiura}, K., {Kubo}, Y., {et~al.} 2017, \apj, 835, 156

\bibitem[{{Park} {et~al.}(2012){Park}, {Cho}, {Bong}, {Kumar}, {Chae}, {Liu},
  \& {Wang}}]{park12}
{Park}, S.-H., {Cho}, K.-S., {Bong}, S.-C., {et~al.} 2012, \apj, 750, 48

\bibitem[{{Park} {et~al.}(2013){Park}, {Kusano}, {Cho}, {Chae}, {Bong},
  {Kumar}, {Park}, {Kim}, \& {Park}}]{park13}
{Park}, S.-H., {Kusano}, K., {Cho}, K.-S., {et~al.} 2013, \apj, 778, 13

\bibitem[{{Park} {et~al.}(2008){Park}, {Lee}, {Choe}, {Chae}, {Jeong}, {Yang},
  {Jing}, \& {Wang}}]{park08}
{Park}, S.-H., {Lee}, J., {Choe}, G.~S., {et~al.} 2008, \apj, 686, 1397

\bibitem[{{Pedregosa} {et~al.}(2012){Pedregosa}, {Varoquaux}, {Gramfort},
  {Michel}, {Thirion}, {Grisel}, {Blondel}, {M{\"u}ller}, {Nothman}, {Louppe},
  {Prettenhofer}, {Weiss}, {Dubourg}, {Vanderplas}, {Passos}, {Cournapeau},
  {Brucher}, {Perrot}, \& {Duchesnay}}]{pedr12}
{Pedregosa}, F., {Varoquaux}, G., {Gramfort}, A., {et~al.} 2012, arXiv
  e-prints, arXiv:1201.0490

\bibitem[{{Pevtsov} {et~al.}(1994){Pevtsov}, {Canfield}, \& {Metcalf}}]{pevt94}
{Pevtsov}, A.~A., {Canfield}, R.~C., \& {Metcalf}, T.~R. 1994, \apjl, 425, L117

\bibitem[{{Priest} \& {Forbes}(2002)}]{prie02}
{Priest}, E.~R. \& {Forbes}, T.~G. 2002, \aapr, 10, 313

\bibitem[{{Scherrer} {et~al.}(2012){Scherrer}, {Schou}, {Bush}, {Kosovichev},
  {Bogart}, {Hoeksema}, {Liu}, {Duvall}, {Zhao}, {Title}, {Schrijver},
  {Tarbell}, \& {Tomczyk}}]{sche12}
{Scherrer}, P.~H., {Schou}, J., {Bush}, R.~I., {et~al.} 2012, \solphys, 275,
  207

\bibitem[{{Schou} {et~al.}(2012){Schou}, {Scherrer}, {Bush}, {Wachter},
  {Couvidat}, {Rabello-Soares}, {Bogart}, {Hoeksema}, {Liu}, {Duvall}, {Akin},
  {Allard}, {Miles}, {Rairden}, {Shine}, {Tarbell}, {Title}, {Wolfson},
  {Elmore}, {Norton}, \& {Tomczyk}}]{scho12}
{Schou}, J., {Scherrer}, P.~H., {Bush}, R.~I., {et~al.} 2012, \solphys, 275,
  229

\bibitem[{{Schrijver}(2007)}]{schr07}
{Schrijver}, C.~J. 2007, \apjl, 655, L117

\bibitem[{{Shibata} \& {Magara}(2011)}]{shib11}
{Shibata}, K. \& {Magara}, T. 2011, Living Reviews in Solar Physics, 8, 6

\bibitem[{{Sinha} {et~al.}(2019){Sinha}, {Srivastava}, \& {Nandy}}]{sinh19}
{Sinha}, S., {Srivastava}, N., \& {Nandy}, D. 2019, \apj, 880, 84

\bibitem[{{Spirock} {et~al.}(2002){Spirock}, {Yurchyshyn}, \& {Wang}}]{spir02}
{Spirock}, T.~J., {Yurchyshyn}, V.~B., \& {Wang}, H. 2002, \apj, 572, 1072

\bibitem[{{Sudol} \& {Harvey}(2005)}]{sudo05}
{Sudol}, J.~J. \& {Harvey}, J.~W. 2005, \apj, 635, 647

\bibitem[{{Tian} {et~al.}(2002){Tian}, {Liu}, \& {Wang}}]{tian02}
{Tian}, L., {Liu}, Y., \& {Wang}, J. 2002, \solphys, 209, 361

\bibitem[{{Tur} \& {Priest}(1976)}]{tur76}
{Tur}, T.~J. \& {Priest}, E.~R. 1976, \solphys, 48, 89

\bibitem[{{van Ballegooijen} \& {Martens}(1989)}]{vanb89}
{van Ballegooijen}, A.~A. \& {Martens}, P.~C.~H. 1989, \apj, 343, 971

\bibitem[{{Vemareddy} {et~al.}(2012){Vemareddy}, {Ambastha}, {Maurya}, \&
  {Chae}}]{vema12}
{Vemareddy}, P., {Ambastha}, A., {Maurya}, R.~A., \& {Chae}, J. 2012, \apj,
  761, 86

\bibitem[{{Wang} \& {Tang}(1993)}]{wang93}
{Wang}, H. \& {Tang}, F. 1993, \apjl, 407, L89

\bibitem[{{Wang} {et~al.}(1996){Wang}, {Shi}, {Wang}, \& {Lue}}]{wang96}
{Wang}, J., {Shi}, Z., {Wang}, H., \& {Lue}, Y. 1996, \apj, 456, 861

\bibitem[{{Welsch} {et~al.}(2009){Welsch}, {Li}, {Schuck}, \&
  {Fisher}}]{wels09}
{Welsch}, B.~T., {Li}, Y., {Schuck}, P.~W., \& {Fisher}, G.~H. 2009, \apj, 705,
  821

\bibitem[{{Yu} {et~al.}(2009){Yu}, {Huang}, {Wang}, \& {Cui}}]{yu09}
{Yu}, D., {Huang}, X., {Wang}, H., \& {Cui}, Y. 2009, \solphys, 255, 91

\bibitem[{{Yuan} {et~al.}(2010){Yuan}, {Shih}, {Jing}, \& {Wang}}]{yuan10}
{Yuan}, Y., {Shih}, F.~Y., {Jing}, J., \& {Wang}, H.-M. 2010, Research in
  Astronomy and Astrophysics, 10, 785

\bibitem[{{Zhang} \& {Bao}(1999)}]{zhan99}
{Zhang}, H. \& {Bao}, S. 1999, \apj, 519, 876

\bibitem[{{Zirin} \& {Wang}(1993)}]{ziri93}
{Zirin}, H. \& {Wang}, H. 1993, \solphys, 144, 37

\end{thebibliography}

%\begin{thebibliography}{99}
%thers S., 2012, Journal of Interesting Stuff, 17, 198
%\end{thebibliography}

%%%%%%%%%%%%%%%%%%%%%%%%%%%%%%%%%%%%%%%%%%%%%%%%%%

\end{document}